\documentclass[rnote]{aa} 

\usepackage{natbib}
\bibpunct{(}{)}{;}{a}{}{,}
\def\lap{\hbox{${_{\displaystyle<}\atop^{\displaystyle\sim}}$}}
\def\gap{\hbox{${_{\displaystyle>}\atop^{\displaystyle\sim}}$}}

\usepackage{txfonts}

\newcommand{\nv}        {\mbox{\boldmath$n_v$}}
\newcommand{\oms}       {\mbox{\boldmath$\omega_s$}}
\newcommand{\omb}       {\mbox{\boldmath$\omega_b$}}

\begin{document}
   \title{Incompatibility of long-period neutron star precession with
creeping neutron vortices}

   \author{B. Link
   \thanks{Permanent address: Montana State
University, Department of Physics, Bozeman, MT 59717, USA}
}

   \institute{Dipartimento di Fisica "Enrico Fermi", 
              Universit\`a di Pisa, 
              Largo Pontecorvo 3, 
              Pisa, I-56127 Italy
\\
              \email{link@physics.montana.edu}
}

   \date{Received May 23, 2006; accepted August 1, 2006}

 
\abstract
  {}
  {To
  determine whether ``vortex creep'' in neutron stars, the slow
  motion of neutron vortices with respect to pinning sites in the core
  or inner crust, is consistent with observations of long-period
  precession.}  {Using the concept of vortex drag, I discuss the
  precession dynamics of a star with imperfectly-pinned ({\sl i.e.},
  ``creeping'') vortices. } {The precession frequency is far too high
  to be consistent with observations, indicating that the standard
  picture of the outer core (superfluid neutrons in co-existence
with type II, 
superconducting protons) should be reconsidered. There is a
  slow precession mode, but it is highly over-damped and cannot
  complete even a single cycle. Moreover, the vortices of the inner crust
must be able to move with little dissipation with respect to the solid.}  {}


\maketitle

There is mounting evidence that some isolated neutron stars undergo
long-period precession (nutation). The strongest
evidence is found in PSR 1828-11, which shows highly-periodic
variations in pulse phase over a period of $\sim 500$ d, accompanied
by correlated changes in beam width 
\citep{sls00}. PSR 1642-03 also shows periodic changes in pulse phase,
though correlated changes in the beam have not been detected
\citep{slu01}. RX J0720.9-3125 is the first
x-ray pulsar to show evidence for precession; the precession period is
$\sim 7$ yr, with correlated changes in line depth
\citep{haberl_etal06}. Low-level timing ``noise'', seen
in all pulsars, is quasi-periodic in many cases, and could represent
precession at low amplitude (for examples see, e.g, 
\citealt{dr83} and \citealt{dalessandro_etal93}). Physically-motivated
models of precession provide good fits to the data of PSR 1828-11
\citep{le01,alw06} and RX J0720.9-3125 \citep{haberl_etal06},
supporting a precession interpretation.
 
The manner in which a neutron star precesses depends on the dynamics
of its interior, and so observations of precession can be used to
study the stellar interior. The purpose of this Note is to discuss the
role that {\em quantized neutron vortices}, which are expected to
occupy most of the neutron star interior, play in the dynamics of
neutron star precession. I focus on recent work on this question and
emphasize that pinning of vortices to magnetic flux tubes in the core,
or to nuclei in the inner crust, is incompatible with observations of
long-period precession.

The first part of this Note is aimed at constraining the vortex
dynamics and state of the outer core. To make this argument, I
assume {\em a priori} that there is no pinning or significant
dissipation associated with the dynamics of vortices in the {\em inner
crust}. At the end of the Note, I argue that the assumption of weak
dissipation in the inner crust is required by observations,
and obtain an upper bound on how strong the dissipation can be.

Nucleon pairing calculations predict that the outer core of a neutron
star consists of a neutron superfluid in coexistence with
superconducting protons (for a review, see, \citealt{dh03}). The
superconductor is expected to be in a type II state \citep{bpp69}, so
that the magnetic field that penetrates the core is organized in flux
tubes \citep{bppr69}, long structures of microscopic cross
section. The flux tubes probably have a very complicated arrangement
that froze to the core medium when it became a superconductor shortly
after the star's formation \citep{rzc98,jones06}. The superconducting
protons, other charges ({\sl e.g.}, electrons and muons) and the crust
are all coupled together through magnetic stresses over time scales of
several seconds, nearly corotating as a rigid body
\citep{easson79}. By contrast, the neutron superfluid rotates by establishing a
nearly rectilinear array of vortex lines, whose arrangement determines
the angular momentum of the neutron fluid.  The neutron vortices of
the outer core, which are themselves magnetized through Fermi liquid
effects
\citep{als84}, {\em pin} to the flux tubes \citep{sauls89}. The origin of the
pinning is that bringing a (magnetized) vortex close to a flux tube
raises (or lowers, depending on orientation) the magnetic energy by
$\sim 5$ MeV per intersection. For typical neutron star rotation rates
and magnetic fields, there are $\gap 10^{16}$ vortex lines and $\sim
10^{31}$ flux tubes. The vortices are thus tangled in the flux tube
array, with an intersection spacing of $\sim 10^{-10}$ cm along a
vortex. This entanglement prevents the neutron fluid from corotating with
the rest of the star. 

From the standpoint of precession dynamics, the star can be regarded
as consisting of two components: 1) the core protons, other charges
and crust, which rotate as a single ``body'', and, 2) a neutron
superfluid which rotates according to the distribution and movement of
its vortices. The angular momentum vector of the superfluid (most of
the star) is fixed to the body to the extent that the vortices remain
pinned against flux tubes, while, in a precessing body, the angular
velocity vector changes its orientation with respect to the body's
symmetry axes. As I argue in this Note, this immobilization of
vortices with respect to the body gives a precession period that is
far to fast to reconcile with observations.

Precession is a rotational mode of a rigid body in which the body's
angular velocity is not aligned with any principal axis. 
Recall that for a rigid, biaxial body of oblateness $\epsilon$, the
precession frequency $\varpi$, in units of the spin frequency
$\omega$, is $\varpi = \epsilon$. If a neutron star precesses
approximately as a rigid body, the observed precession period of PSR
1828-11 implies $\epsilon\simeq 10^{-8}$; the other precession
candidates imply similar values of $\epsilon$.

Pinning of vortices anywhere dramatically increases the precession
frequency. Shaham (1977) showed that if vortex pinning is perfectly
effective then precession would occur at very high frequency: 
\begin{equation}
\varpi = \epsilon + \frac{L_p}{I_b\omega}, 
\end{equation}
where $L_p$ is the angular momentum contained in the pinned superfluid
and $I_b$ is the moment of inertia of the ``body'' (crust plus
charges). The pinned fluid effectively acts as a huge additional
contribution to the star's oblateness, since $L_p\simeq I_p\omega$ and
$L_p/I_b\omega\simeq I_p/I_b>>\epsilon$. The precession frequency is
thus determined by how much of the neutron vorticity is pinned. If
most of the neutron superfluid of the outer core is pinned to the flux
tubes, the precession frequency will be $\varpi\simeq I_p/I_b\simeq
10$, compared to $10^{-8}$ observed for PSR 1828-11. Hence, perfect
pinning in even a small fraction of the core gives a precession
frequency that is far too fast to be consistent with
observations. This result is a consequence of the fact that the vortex
array carries enormous vorticity; with the added constraint that
there is vorticity fixed to the body, the star must precess at a
frequency higher than $\epsilon$ in order to conserve angular
momentum. For
precession to occur with long period, the vortices must be able to
closely follow the instantaneous spin vector of the rest of the star,
and so cannot be perfectly pinned in any sizeable region of the star;
the picture of vortex lines entangled in flux tubes appears to be
incompatible with observations of long-period precession
\citep{link03}.

The assumption of perfect pinning is, however, an idealization. Some
vortex motion with respect to pinning sites could occur by a
process of {\em creep} through thermal activation 
\citep{alpar_etal84,leb93,cc93} or quantum tunneling
\citep{leb93}. \citet{alpar05} has suggested that vortex creep could
resolve the problem that precession is too fast with perfectly-pinned
vortices. I now show that even if core vortices creep with respect to
the flux tubes against which they are pinned, the precession frequency
is still too fast, by a factor of $\sim 10^9$, to explain the
observations.

The problem at hand is how the creep of vortices with respect to
defects affects precession. The defects I will first consider are flux
tube/vortex junctions in the outer core. I will then consider the
nuclei of the inner crust.  The formulation of Sedrakian, Wasserman
\& Cordes (1999; hereafter SWC), who studied precession of a star with
neutron vortices using a description of vortex motion based on the
concept of vortex drag, is particularly useful for addressing this
question. They introduced a drag force per unit length of vortex of
the form (SWC, eq. 29, with notational changes)
\begin{equation}
{\mathbf f_d} = -\eta({\mathbf v_v}-{\mathbf v_b}) 
- \eta^\prime \nv\times({\mathbf v_v}-{\mathbf v_b}), 
\label{fd}
\end{equation}
where ${\mathbf v_v}$ is the velocity of a vortex line, ${\mathbf
v_b}$ is the velocity of the background body against which the
vortices are dragged, $\eta$ and $\eta^\prime$ are drag coefficients,
and $\nv$ is a unit vector in the direction of the angular velocity of
the superfluid (coincident with the vorticity axis and also parallel
to the vortex array). In the outer core, the background against
which the vortices are dragged is the fence of flux tubes. The first
term refers to (dissipative) drag anti-parallel to the vortex motion,
while the second term describes a (possible) non-dissipative force
transverse to the vortex motion. It is convenient to define the
dimensionless drag coefficients
\begin{equation}
\beta = \frac{\eta\rho_s\kappa}{(\rho_s\kappa-\eta^\prime)^2+\eta^2},
\label{eta}
\end{equation}
\begin{equation}
\beta^\prime=1-\frac{\rho_s\kappa(\rho_s\kappa-\eta^\prime)}
{(\rho_s\kappa-\eta^\prime)^2+\eta^2},
\label{etaprime}
\end{equation}
where $\rho_s$ is the superfluid mass density and $\kappa$ ($\equiv
h/2m_n$; $m_n$ is the neutron mass) is the
quantum of vorticity in a neutron superfluid. 
The drag coefficients determine how the vortex moves with respect to
the background (SWC, eq. 33):
\begin{equation}
{\mathbf v_v} - {\mathbf v_b} = (\beta^\prime - 1)({\mathbf v_b} - 
{\mathbf v_s}) + 
\beta \nv\times ({\mathbf v_b} - {\mathbf v_s}), 
\label{vv}
\end{equation}
where ${\mathbf v_s}$ is the velocity of the neutron superfluid. 

Vortex creep, by definition, means that ${\mathbf v_v}\simeq {\mathbf
v_b}$, which from eqs. (\ref{eta})--(\ref{vv}), implies that
$\eta/\rho_s\kappa>>1$. {\em Hence, vortex creep is in the high-drag
limit}. To calculate the precession dynamics, however, we must know
the relative magnitudes of $\eta$ and $\eta^\prime$. The transverse
force (second term of eq. \ref{fd}) has not been shown to exist in
any calculation of drag on vortices by any process in a neutron star. 
The existence of this term is, in fact, a controversial issue
in laboratory superfluids. On the one hand, \citet{thouless_etal96}
and \citet{wexler97} have argued that this term is not present at all
for an isolated vortex moving in an infinite, uniform superfluid (but
see the comments by
\citealt{hall_hook98} and \citealt{sonin98} and the replies
therein). On the other hand, mutual friction experiments with
superfluid He (for which the system is, of course, neither infinite
nor uniform) do demonstrate the existence of the transverse force
({\sl e.g.}, \citealt{bevan_etal97}). Whether a transverse force
exists for a vortex moving with respect to defects in a neutron star
is unknown, so I will consider the possibility that it exists to
maintain generality and appeal to laboratory experiments to obtain a
bound. All mutual friction experiments in superfluid He that measure
$\eta^\prime$ find that this coefficient goes to zero for temperatures
well below the superfluid critical temperature ({\sl e.g.},
\citealt{bevan_etal97}), the appropriate temperature regime for a
neutron star. As the temperature is increased,
$\eta^\prime/\rho_s\kappa$ usually increases, but never exceeds
unity. Unless $\eta^\prime$ has very different behavior in a neutron
star, then $\eta>>\eta^\prime$ in the limit of high drag. I henceforth
assume this, but will return to the unlikely possibility of
$\eta^\prime>>\eta$ at the end of this Note. Eqs. (\ref{eta}) and
(\ref{etaprime}) become
\begin{equation}
\beta\simeq\frac{\eta\rho_s\kappa}{\eta^2+(\rho_s\kappa)^2} \qquad\qquad
\beta^\prime \simeq \frac{\eta^2}{\eta^2+(\rho_s\kappa)^2}. 
\label{coeffs}
\end{equation}
If there is no drag
($\eta=\beta=\beta^\prime=0$, with $\eta^\prime=0$), then ${\mathbf v_v}={\mathbf v_s}$ and the
vortices corotate with the superfluid. When the vortices are dragged,
${\mathbf v_s}$ and ${\mathbf v_v}$ differ. Perfect pinning
corresponds to $\eta\rightarrow\infty$, $\beta=0$, $\beta^\prime=1$.
In the high-drag limit ($\eta/\rho_s\kappa>>1$),
\begin{equation}
\beta \simeq \frac{\rho_s\kappa}{\eta}<<1 \qquad\qquad 1-\beta^\prime
\simeq\beta^2.
\end{equation}
SWC considered a two-component biaxial star with vortices moving
according to the drag force of eq. (\ref{fd}), and obtained the modes of the
system. In the limit of high drag, the limit appropriate to vortex
creep, the modes take simple forms. 
There is a high-frequency mode of complex frequency (SWC, eq. 81, with
$\varpi_f\equiv -ip_p$) 
\begin{equation}
\varpi_f = \frac{I_p}{I_b} + i
\beta\left(1+\frac{I_p}{I_b}\right). 
\label{fast}
\end{equation}
The real part gives the precession frequency, and the imaginary part
gives the damping rate.  This mode corresponds to precession at very
high frequency $(I_p/I_b)\omega$, as found by Shaham (1977),
with little damping (since $\beta<<1$). But there is also a new slow
mode, of frequency (SWC, eq. 80, with $\varpi_s\equiv -ip_d$)
\begin{equation}
\varpi_s = -\epsilon\frac{I_b}{I_p}\beta^2+i\epsilon\frac{I_b}{I_p}\beta.
\end{equation}
This mode can have a very low frequency compared to $\omega$. The mode
is absent (has zero frequency) in the case of perfect pinning; it
arises only if the vortices can move with respect to the body. Since
$\beta<<1$ in the high-drag limit, as relevant to vortex creep, the
damping rate of the mode is faster than the oscillation frequency by a
factor of $\beta^{-1}>>1$. Hence, this mode is not oscillatory at all;
{\em it is highly over-damped, and never completes even one
cycle}. The problem becomes even more serious the more effective the
pinning is (smaller $\beta$). {\em Hence even if the vortices move
through creep}, the fast precession mode of eq. (\ref{fast}) is the
only persistent and dynamically relevant one. The effects of
triaxiality do not change the basic picture (SWC).

\citet{alpar05} concluded that the slow mode is under-damped
based on an assumed torque between the crust and the superfluid of the
following linear form:
\begin{equation}
{\mathbf N_{\rm drag}} = -(\oms - \omb)/\tau
\label{alpar_torque}
\end{equation}
where $\tau$ is a damping time, $\oms$ is the angular velocity of the
superfluid and $\omb$ is the angular velocity of the body.  This drag
torque does not correctly describe the dynamics of a dragged vortex
array; the correct form, obtained by integrating ${\mathbf
r}\times {\mathbf f_d}$ from eq. (\ref{fd}) throughout the star, takes a
more complicated form (see SWC, eqs. 50 and 51), and is much larger
than that of eq. (\ref{alpar_torque}). Use of eq. (\ref{alpar_torque})
greatly underestimates the strength of the damping that arises from
dissipative vortex motion, and leads to the incorrect conclusion that
the slow precession mode is under-damped.

Provided the assumption that the vortices of the crust are not
significantly dragged remains valid, a simple and general conclusion
follows from this analysis: long-period precession demands that the
vortices of the outer core move with little drag, that is, they do not
creep, but {\em flow} essentially everywhere. Hence, in the outer
core, {\em vortices and flux tubes cannot co-exist}
\citep{link03}. The possibilities that
might be consistent with long-period precession are: 1) normal
neutrons, superconducting protons, 2) superfluid neutrons, type I
protons, 3) superfluid neutrons, normal protons, 4) normal neutrons
and protons, and, 5) the core magnetic field has somehow been almost
completely expelled.  The existence of proton superconductivity is on
rather firm footing, though the {\em type} of superconductivity is
less clear.  In a type I core, the magnetic flux could be organized in
slabs or other geometries of mesoscopic dimensions, rather than the
flux tubes found in a type II superconductor.  It might be possible
for vortices to move through a type I superconductor with sufficiently
low drag to allow long-period, under-damped precession;
\citet{sedrakian05} showed that this scenario works for two specific
geometries for the magnetic field. The pairing state of neutrons is
far less certain. Most pairing calculations show pairing, but at least
one calculation does not
\citep{sf04}. Lacking a mechanism for complete expulsion of the core
field, possibilities 1) and 2) seem the most likely within existing
uncertainties regarding nucleon pairing. Until the issue of neutron
pairing in the outer core is settled by first-principles calculations,
astrophysical arguments of the type presented here are useful for
constraining the various possibilities for the hadronic ground
states. 

I now turn to the inner crust, where vortices could pin to lattice nuclei
\citep{alpar77,eb88,pvb97,abbv06,dp06}. At the relatively low densities
of the inner crust, neutron pairing is well-understood and neutron
superfluidity there is theoretically well-established (see 
\citealt{dh03} for a review).
Assuming, for example, that almost all of the inner-crust superfluid
is pinned to nuclei and that the rest of the star rotates as a rigid
body over time scales less than the precession period, the precession
frequency becomes $\varpi\simeq 0.01$ \citep{shaham77}, again far too
fast. There is a slow mode if the vortices creep, but as before it is
highly over-damped. This result does not necessarily rule out vortex
pinning in the inner crust; it just cannot happen in stars that are
slowly precessing. In fact, the hydrodynamic forces acting on {\em
pinned vortices} in a star with a ``wobble angle'' $\theta$ between
the star's symmetry axis and the angular momentum of $\sim 3^\circ$,
as inferred for PSR 1828-11, would be sufficient to unpin the vortices
in the inner crust \citep{lc02}. The question then becomes one of how
strong these unpinned vortices are dragged. For the precession to be
of long period, the vorticity axis must be able to closely follow the
instantaneous rotation axis of the body. To do this, the vortices must
move at a speed
\begin{equation}
\vert {\mathbf v_v} - {\mathbf v_b}\vert \simeq R\omega\,\theta\simeq
10^{-2}
\mbox{ cm s$^{-1}$   (PSR 1828-11)}. 
\end{equation}
Here $R$ is the stellar radius and $\theta$ is the wobble angle. The
motion of vortices past the nuclei with which they interact excites
waves on the vortices (Kelvin modes), an inherently dissipative
process. \citet{eb92} and \citet{jones92} studied this problem in the
context of explaining the short spin-up time scale of pulsar
glitches. They found Kelvin mode excitation to be highly dissipative
for $\vert {\mathbf v_v} - {\mathbf v_b}\vert\gap 10^6$ cm s$^{-1}$,
a characteristic velocity difference that could develop as a consequence
of vortex pinning. These treatments were done in a perturbative
approximation, cannot be extended below $\vert {\mathbf v_v} -
{\mathbf v_b}\vert\sim 10^5$ cm s$^{-1}$, and so cannot be applied to
the precession problem; instead we require a description of vortex
drag at much lower velocities, in which the vortex response to motion
past a nucleus is highly non-linear. \citet{jones98} showed that an
infinitely long vortex moving slowly past a single nucleus tends to be
temporarily trapped by the nucleus, and argued that vortices will pin,
making it theoretically impossible (at least, within our current
understanding of the inner crust) for a neutron star to precess slowly
\citep{jones04}. The treatment of \citet{jones98}, however, gives an unphysical divergence
in the dissipation rate at low velocities when the rate should instead
vanish \citep{link04}. For vortex dynamics in the inner crust to allow
long-period precession, assuming no difficulties with the core of the
sort described above, the drag coefficient must satisfy
$\eta/\rho_s\kappa << I_b/I_s$ \citep{link04,sedrakian05}; here $I_s$
is the moment of inertia of the inner crust superfluid and $I_b$ is
now the moment of inertia of the crust plus all other components
tightly coupled to it over time scales less than the precession
period. If $I_b$ comprises most of the star, then $I_b/I_s\simeq 100$,
which allows some room for vortex motion at moderately strong drag,
$\eta/\rho_s\kappa\sim 1-10$ for example. Whether or not $\eta$ can
satisfy this upper bound for $\vert {\mathbf v_v} - {\mathbf v_b}\vert
\simeq 10^{-2}$ is an interesting, unsolved problem.

Based on the above arguments, I conclude that in slowly precessing
neutron stars: 
\begin{enumerate}
\item vortices do not pin essentially anywhere in the star, so that, 
\item the standard picture of the neutron star outer core in which
vortices coexist with flux tubes should be reconsidered, and, 
\item inner crust vortices must move with little
dissipation ($\eta/\rho_s\kappa\lap 10$) with respect to the solid. 
\end{enumerate}
These results are robust provided that precession is real; they are not
compromised by the numerous uncertainties regarding the composition
and dynamics of the inner core. The only possible loophole for vortex
creep to be compatible with long-period precession seems to be that
$\eta^\prime$ is not only finite, but for some reason satisfies
$\eta^\prime/\rho_s\kappa>>\eta/\rho_s\kappa>>1$ everywhere that
vortices are pinned. Only under these
unlikely circumstances can vortex creep admit a long-period,
under-damped precession mode (SWC).

\acknowledgements
I thank A. Sedrakian and I. Wasserman for valuable discussions, and
the referee, P. B. Jones, for extremely helpful criticisms. Much of the
work described here was supported by U.S. NSF grant AST-0406832.

\end{document}